\documentstyle[aps,preprint,epsfig]{revtex} 
\begin{document} 
\tightenlines 
 
\title{Setting Confidence Belts} 
\author{Byron P. Roe and Michael B. Woodroofe}  
\address {Department of Physics (B.P.R) and Department of Statistics
(M.B.W) University of Michigan, Ann Arbor, MI 48109}  
 
\maketitle 
 
\begin{abstract} 
We propose using a Bayes procedure with uniform improper prior to 
determine credible belts for the mean of a Poisson distribution in the 
presence of background and for the continuous problem of measuring a 
non-negative  
quantity $\theta$ with a normally distributed measurement error.  
Within the Bayesian framework, these belts are optimal.   
The credible limits are then examined from a frequentist point of view 
and found to have good frequentist and conditional frequentist 
properties. 
\end{abstract} 

\section{Introduction} 
We consider two simple problems for setting confidence belts. The 
first problem is that of the Poisson distribution in the presence  
of background.  The second problem is the 
measurement of a non-negative parameter $\theta$, with a normally
distributed error.  
Although simple problems, there has been considerable work and 
discussion on them in the last few years.   
Most of the discussion has centered on the Poisson case, when the 
number of observed events is small. 
This is an important current topic because there are  many search 
experiments involving small numbers of events.  These include 
Higgs particle searches, supersymmetry searches, and neutrino 
oscillation searches such as LSND and KARMEN.   
 
We will briefly describe some recent attempts to address these 
problems.   
 
\section{The Feldman-Cousins Unified Approach} 
The standard method for setting 90\% CL bounds has been to select a 
region with a 5\% probability above and a 5\% probability below the 
bounds.  About two years ago, 
Gary Feldman and Bob Cousins\cite{coustwo} 
suggested a method, new to physics 
analyses, called the unified procedure.  For the Poisson 
case, for each 
possible value of the parameter $\theta$, they looked at the ratio of 
the probability of getting the observed number of events $n$ for  
that $\theta$ compared to 
the maximum probability for  all physically allowed $\theta$'s, and
picked $n$'s with the highest ratio to build a  
90\% confidence region.   
This solved two problems with the old procedure.  It automatically 
transitioned from an upper limit to a confidence belt and it always 
produced confidence sets in the physical region.   
 
The unified procedure works well for many problems and is a 
significant improvement over the symmetric tails procedure.  However 
it has a serious problem if few or no events are observed.  If 0 
events are seen, there are 0 signal and 0 background events. 
That there are 0 background events is interesting, but irrelevant to
the question of whether signal  
events are seen.  A 90\% C.L. limit on $\theta$ should come from 
$p_\theta(0)\le 10$\%, which sets a limit at $\theta=2.3$. 
(Throughout this article probabilities for a fixed 
parameter $\lambda$ are denoted by a subscript $p_\lambda$.  Probability 
densities and probability masses are denoted with lower case 
letters and distribution functions by upper case letters.) 
Such a case 
did occur in the Summer of 1998 from initial KARMEN  
results \cite{karmen}.   They had 
0 events with a background of 2.88, and using the unified method obtained 
an upper limit of $\theta = 1.08$.  It is desirable to have a method 
set a confidence limit near 2.3, independent of background, for a 
90\% C.L., when no events are seen. 
 
\section{The Roe-Woodroofe Procedure} 
We \cite{roewood} presented a variant of the Feldman-Cousins unified 
procedure which corrected this problem and  
introduced the concept of conditional coverage. For any observation we 
know that the background is less or equal to the observed number of events, 
$n_{observed}$.   
We suggested the use of 
a sample space not of all experimental outcomes, but of outcomes with  
background  
$\le n_{observed}$.  Thus, if 
$n_{observed}=4$, we considered only measurements in which there are 
$\le 4$ background events, even for 20 events observed in the sample 
space.  This probability  
was used both to set the ratio 
used in the unified method and to calculate the coverage probability. 
This method had the advantage of giving an upper limit of about 2.42 
for 0 observed events independently of the background mean $b$.   
See Figure 1.   
 
Robert Cousins \cite{cousthree} found that 
this procedure had a problem with its lower limit when applied to a 
continuous observable.  Suppose one measures a parameter $\theta \ge 0$  
with a measurement 
error $\Delta,\ x=\theta+\Delta$, where $\Delta$ is normal (0,1).   
The R-W procedure eliminated  
$\theta=0$ for all $x>0$.  However, if $\theta=0$, the probability of 
measuring $x>1.39$ is 10\%.  See Figure 2. 
It appeared that the R-W upper limit and the F-C lower limit were needed. 
 
Since the conditional coverage concept is, as yet, unfamiliar, it is
desirable   
that a method should have reasonable conventional coverage as well as 
conditional coverage. 
 
\section{Bayesian Credible Intervals} 
 
In this section, we derive Baysian credible intervals when the
expected signal is given a uniform prior distribution. These
intervals have several desirable properties. Like the unified
intervals, Bayesian intervals lie in the physical region and
automatically change from confidence bounds to two-sided
intervals. They avoid the known pitfalls mentioned above, the
problems with low counts in unified method and the large lower
boundary of the conditional confidence intervals in the
continuous case. Finally, and importantly, the Bayesian
intervals are optimal on their own terms. The intervals
derived minimize length among all credible intervals. The
price paid for these desirable properties is dependence on the
prior and metric. We address the first of these concerns by
showing that the frequentist coverage probability of the
Bayesian intervals is quite close to the Bayesian posterior
credible level. That is, while the two probabilities are
conceptually quite different, they are close numerically. In
addition, we show that the Bayesian credible intervals have
exact conditional (frequentist) coverage probability, except
for discreteness in the Poisson case. With regard to the
metric, it is primarily the optimality of our procedures that
depends on the metric. See Section 7.
 
	We will illustrate with two examples, a Poisson 
case in which the mean is composed of an unknown signal mean 
$\theta\ge 0$ and a known background mean $b$ and the measurement $x$  
of a parameter $\theta\ge 0$, with a measurement error $\Delta$ which
is normal (0,1).  We use a uniform (improper) prior of 1 for
$0\le\theta<\infty$ in both problems.

\subsection{The Continuous Example} 
This is a simple problem in which the statistical issues are clear and 
which approximates the Poisson problem for large $b+\theta$. 
We measure $x=\theta + 
\Delta$, where $\theta\ge 0$, and the density function for $x$ is 
\begin{equation} 
f(x|\theta) \equiv \phi(x-\theta) = { 1 \over \sqrt{ 2\pi}} 
e^{-{1\over 2}(x-\theta)^2}.  
\end{equation} 
For the uniform prior probability $pr(\theta )=1$, the marginal  
density of $x$ becomes  
$$ 
f(x) =\int_0^\infty \phi(x-\theta) pr (\theta)d\theta=\Phi(x), 
$$ 
where 
$$\Phi(x)\equiv \int_{-\infty}^x\phi(y)dy,$$ 
the standard normal distribution function.  We now use Bayes theorem 
$f(x|\theta)\times pr(\theta) = f(\theta|x)\times f(x)$.  The  
conditional density 
of $\theta$ given $x$ is 
\begin{equation} 
f(\theta|x) = {\phi(x-\theta) \over \Phi(x)}. 
\end{equation} 
$f(\theta|x)$ is proper; the improper (infinite) prior has 
cancelled out.   
We wish to find an upper limit $u$ and a lower 
limit $\ell$, dependent on $x$, for which 
\begin{equation} 
Prob\{\ell\le\theta\le u|x\} = 1-\epsilon. \label{eq:cred} 
\end{equation} 
In Bayes theory, such intervals are called {\it credible intervals}. 
It is desirable to minimize the interval $[\ell,u]$, subject to
(\ref{eq:cred}).  We do that by 
picking $\theta$'s with the largest probability density to be within
the interval,  
$$[\ell,u] =\{\theta:f(\theta|x)\ge c\},$$ 
where $0\le c\le 1/\sqrt{2\pi}$ is chosen to satisfy Equation (\ref{eq:cred}). 
Now, $f(\theta|x)\ge c$ if and only if $|\theta -x|\le 
d$, where 
$$d = \sqrt{-2\ln c-\ln(2\pi) -2\ln[\Phi(x)]}.$$ 
 
There are two cases to be considered. If $d\le x$,  then the condition
(\ref{eq:cred}) becomes  
$$ 
1-\epsilon = \int_{x-d}^{x+d}f(\theta|x)d\theta =\int_{-d}^d  
{\phi(\omega) \over \Phi(x) } d\omega = { 2\Phi(d) -1 \over 
\Phi(x)};  
$$ 
\begin{equation} 
d = \Phi^{-1}[ {1 \over 2} + {1 \over 2}(1-\epsilon)\Phi(x)], 
\end{equation} 
where $\Phi^{-1}$ denotes the inverse function to $\Phi$.  If $x < d$,
(\ref{eq:cred}) becomes  
$$\epsilon = \int_{x+d}^\infty f(\theta|x)d\theta =\int_d^\infty 
{\phi(\omega) \over \Phi(x) }d\omega = {1-\Phi(d) \over \Phi(x)} ;$$ 
\begin{equation} 
d = \Phi^{-1}[1-\epsilon\Phi(x)]. 
\end{equation} 
 
If $x_0$ is the point where these two curves meet, then, 
\begin{equation} 
x_0 = \Phi^{-1}\left( {1 \over 1+\epsilon}\right). 
\end{equation} 
Hence, the desired credible interval is  
$[\ell,u] = [\max(x-d,0),x+d],$ where 
$d = \Phi^{-1}[1-\epsilon\Phi(x)]$, if  
$ -\infty < x \le x_0$,  
and $d= \Phi^{-1}[{{ 1 \over 2}} + {{ 1 \over 2}}(1-\epsilon)\Phi(x)]$  
if $ x_0<x<\infty$.

\subsection{Poisson Example}   
The probability mass function and distribution function for the  
Poisson distribution with mean $\lambda$ are: 
\begin{equation} 
p_\lambda(n) = { e^{-\lambda}\lambda^n \over n!}. 
\end{equation}  
and $P_\lambda(n) = p_\lambda(0)+\cdots + p_\lambda(n)$.  For our  
present problem $\lambda = \theta + b$, where $b\ge 0$ is the fixed (known) 
``background'' mean.   
If we let $\theta$ have a prior uniform distribution $pr (\theta)=1$  
for $\theta\ge 0$, 
we then have 
$$ 
p(n)=\int_0^\infty{1\over n!}(\theta+b)^ne^{-(\theta+b)} pr (\theta)d\theta  
$$ 
\begin{equation} 
= \int_b^\infty {1 \over n! }\omega^ne^{-\omega}d\omega =P_b(n) 
\end{equation}  
as derived in our previous paper\cite{roewood}.  The derivation 
proceeded by expanding $(\theta+b)^n$ and noting that 
$\int_0^\infty \theta^ke^{-\theta}d\theta=k!$. 
We again use Bayes' Theorem $p_{\theta+b}(n)\times 
pr (\theta) = p_b(n,\theta) = p_b(\theta|n)\times p(n)$.  Then, recalling 
that $pr (\theta)=1$ for $\theta\ge 0$ and $p(n) = P_b(n)$, we have 
\begin{equation}  
p_b(\theta|n) = { p_{\theta+b}(n) \over P_b(n)}. 
\end{equation} 
 
To set upper and lower confidence bounds for $\theta$, we need to 
solve the equation 
\begin{equation} 
1-\epsilon = \int_\ell^up_b(\theta|n)d\theta = { P_{b+\ell_n}(n) - 
P_{b+u_n}(n) \over P_b(n)} 
\end{equation} 
for $\ell$ and $u$.  A 90\% C.L. 
corresponds to setting $\epsilon = 0.1$. 
The second equality here is obtained by writing $\int_\ell^u
=\int_\ell^\infty -\int_u^\infty$ and using the reasoning described
after Equation (8).  To get the shortest 
interval $u-\ell$ we need to include values of $\theta$ with highest 
density, {\it i.e.},  
\begin{equation} 
[\ell,u]=[\theta:p_b(\theta|n)\ge c] 
\end{equation} 
for some $c$.  The conditions (10) and (11) may be solved 
numerically by an iterative procedure.  This procedure is described 
in Section {\bf 6} in a more general context. 
 
\section{Frequentist Properties} 

The Bayesian credible intervals just derived have exact
posterior coverage probability $1-\epsilon$, by construction.
In this section, we show that they also have high frequentist
coverage probability.  
 
\subsection{The Continuous Example} 
 
In the continuous example, $x = \theta + \Delta, \theta \ge 0$,
and $\Delta \sim {\rm normal}(0,1)$.  Fix a $\theta$ and find the
points at this fixed $\theta$ where $x$ meets the 
upper and lower Bayesian limits.  Let $x_\ell(\theta)$ be that $x$ which meets 
the upper limit, $\theta = u(x_\ell) = x_\ell + d(x_\ell)$ and 
$x_u(\theta)$ be that $x$ which meets the lower limit,
$\theta=\ell(x_u) = x_u - 
min[d(x_u),x_u]$.  Then $[\ell(x),u(x)]$ covers $\theta$ iff 
$x_\ell\le x\le x_u$, and the conventional unconditional 
coverage is $\Phi(x_u-\theta) -\Phi(x_\ell-\theta)$.  This is not 
exactly $1-\epsilon$.  A lower limit on the 
coverage is shown in the Appendix to be  
\begin{equation} 
\Phi(x_u-\theta)-\Phi(x_\ell-\theta) \ge {1-\epsilon \over 
1+\epsilon }, 
\end{equation} 
but this is a very conservative limit.  For $\epsilon=0.1$, this lower 
limit is 0.8182, while a numerical calculation finds a minimum 
coverage of about 0.86. 
 
The conventional coverage probability can be improved for a  
very small increase in limits.    We will consider a conservative 
ad-hoc modification of the upper 
limit on $\theta$ using the one-sided limit for a  
higher confidence level, $\epsilon'=\epsilon/2$.  Let  
$$ 
u'(x) = \max\left[u(x),x+\Phi^{-1}\left(1- {1 \over 
2}\epsilon\right)\right].  
$$ 
Let $x'_\ell$ be the $x$ corresponding to $u'(x) = \theta$.   
The formula for the conventional coverage is 
$\Phi(x_u-\theta)-\Phi(x'_\ell-\theta)$, as above.  
The undercoverage, derived in the Appendix, is very small for  
this conservative modification.  For a 90\% C.L., the conventional  
coverage is at least .900 everywhere to three significant figures. 
Figure 3 shows a plot of the coverage bands, plotted as $\theta-x$ 
vs $x$.   
The confidence bands are shown in Figure 4 and compared with the 
old R-W upper bound and the Feldman-Cousins unified procedure lower 
bound.  The conventional frequentist coverage for the 
Bayesian model and for this conservative modification are shown 
in Figure 5. 
 
The Bayesian credible intervals also have an exact conditional
frequentist property, in terms of the error $\Delta=x-\theta$.  Note
that if $x$ is observed, then, necessarily, 
$\Delta\le x$, since $\theta\ge 0$.  $\Phi(x)$ is just the probability that 
$\Delta\le x$.  Let $a(x) = min[d(x),x]$.   
Then the interval $[\ell(x),u(x)]$ covers $\theta$ iff $\ell(x)\le
\theta\le u(x)$ or, 
equivalently, $-d(x)\le\Delta\le a(x)$.  In the Appendix, it is shown 
that if $x'=\theta+\Delta'$ is an independent copy of $x$, then 
$$Prob_\theta[-d(x)\le\Delta'\le a(x)|\Delta'\le x]=1-\epsilon$$ 
and  
$[\ell(x),u(x)]$ is the shortest interval with these properties.

\subsection{The Poisson Example} 
 
As in the continuous case, the {conventional coverage} is not exact in
the Poisson case.  
For $b=3$ and $1-\epsilon = 0.9$, for example, the conventional  
coverage varies from about 86\% to 96.6\%.   
Figure 6 shows the resulting confidence belt for $b=3,\ \epsilon=0.1$. 
  
The conditional coverage is shown in the Appendix to be   
$\approx 1 -\epsilon$, except for discreteness.   

The conventional coverage can be improved by a small ad hoc 
modification similar to that used for the continuous case.  Consider 
an alternate Bayesian upper limit $u'$ for $\theta$ defined as the 
one-sided limit for a credible level $ = 1-\epsilon'$.  Take 
the modified upper limit as the maximum of $u$ and $u'$.  In the 
continuous example we chose $\epsilon'=\epsilon/2$.  Here, we make a
different ad hoc choice.  The effect of 
the modification for $b = 3,\ \epsilon=0.1$ and $\epsilon'=0.08$ 
is shown in Figure 7.  The value of 
the limit for $n=0$ then increases from 2.3 to 
2.53.  See Figure 8.  For a fixed $\epsilon'$ the limit at $n=0$ is  
independent of $b$, but the appropriate choice of $\epsilon'$ will 
depend weakly on $b$. 
   
\section{Partial Background-Signal Separation} 
 
In this section, we extend the method to processes in which we have
some information concerning 
whether a given event is a signal or a background event.   
Suppose on each event one measures a statistic $x$ (or a vector of 
statistics) for which the density function is $g(x)$ for signal events 
and $h(x)$ for background events.  Suppose, further, that the  
number of observed 
events $n$ has a Poisson distribution with mean $b+\theta$, where
$\theta$ is the expected number of signal events.  We assume that $b$,
the expected  
number of background events, is known.  Then the joint probability 
mass function/density for observing $n$ events and parameters 
$x_1,\cdots,x_n$ is 
$$ 
f_\theta(n,x) = {(b+\theta)^n \over n!}e^{-(b+\theta)} 
\prod_{k=1}^n \left[{\theta g(x_k) +bh(x_k) \over b+\theta}\right]  
$$ 
\begin{equation} 
={1 \over n!}e^{-(b+\theta)}\prod_{k=1}^n[b +\theta r(x_k)]\times  
\prod_{k=1}^n h(x_k), 
\end{equation} 
where $r(x_k) = g(x_k)/h(x_k)$.  For a uniform prior, $pr (\theta)=1$ for 
$0\le\theta < \infty$, the marginal and posterior probability mass
function/densities of 
$n$ and $x_1,\cdots,x_n$, and of $\theta$ are then 
\begin{eqnarray*} 
f(n,x) &=& \int_0^\infty f_\theta(n,x)d\theta \\ 
&=& { 1 \over
n!}\prod_{k=1}^nh(x_k)\times\int_0^\infty
e^{-(b+\theta)}  
\prod_{k=1}^n [b +\theta r(x_k)]e^{-(b+\theta)}d\theta  
\end{eqnarray*} 
 
\begin{equation} 
q(\theta|n,x) = {f_\theta(n,x) \over f(n,x)}. 
\end{equation} 
Observe that 
$$ 
\int_0^\infty e^{-(b+\theta)}\prod_{k=1}^n [b +\theta 
r(x_k)] d\theta =n!\sum_{m=0}^nC_{n,m}{b^{n-m} \over 
(n-m)!} e^{-b} , 
$$ 
where 
\begin{equation} 
C_{n,m} = \left[1/\left({ n \atop m}\right)\right] \sum_{j_1 +\cdots 
+ j_n=m}\prod_{k=1}^n r(x_k)^{j_k},  
\end{equation} 
and where $j_1\cdots j_n$ are restricted to the values 0 or 1.  Cancelling 
common factors, 
\begin{equation} 
q(\theta|n,x) = {e^{-\theta}\prod_{k=1}^n[b+\theta r(x_k)] \over 
n!\sum_{m=0}^n C_{n,m} { b^{n-m} \over (n-m)!}}. \label{eq:q} 
\end{equation} 
Observe also that if $g=h$, then $C_{n,m}=1$ and $q(\theta|n,x)$ is the same 
as the posterior density of $\theta$ given $n$ in the Poisson case 
without the extra variables $x$. 
 
We want to find upper and lower limits $u,\ell$ for $\theta$ such that 
$Prob\{ \ell\le\theta\le u|n,x\} = 1-\epsilon$ and to minimize that  
interval.  This means we want $[\ell,u]=\{\theta:q(\theta|n,x)\ge c\}$. 
We first find the value $\theta_{max}$ at 
which $q(\theta|n,x)$ is maximum, or, equivalently, $\ln 
[q(\theta|n,x)]$ is maximum.  Since the denominator in Equation (\ref{eq:q}) 
\begin{equation} 
D = n!\sum_{m=0}^n {b^{n-m} \over (n-m)!}C_{n,m} \label{eq:d} 
\end{equation} 
does not depend on $\theta$, 
$$ 
{d \over d\theta} \ln[q(\theta|n,x)] = \sum_{k=1}^n {r(x_k) \over 
b+\theta r(x_k) } -1. 
$$ 
If $r(x_1)\cdots r(x_n)<b$, then $\theta_{max}=0$. Otherwise,
setting the derivative equal to 0 then leads to the equation 
\begin{equation} 
\theta_{max} = \sum_{k=1}^n {\theta_{max} r(x_k) \over b +
\theta_{max}  \label{eq:thetahat} 
r(x_k)}.  
\end{equation} 
If $g=h$, then it is easy to see that $\theta_{max}= \max[0,n-b]$. 
It is clear from (\ref{eq:thetahat}) that $\theta_{max}\le n$, and it
can be shown that 
the obvious iteration starting at $\theta=n$ converges to
$\theta_{max}$.  This 
iteration is a special case of the EM algorithm \cite{emalg}. 
 
Next integrate $q(\theta|n,x)$. 
Let  
$$ 
Q(a|n,x) \equiv \int_0^aq(\theta|n,x)d\theta. 
$$ 
Then  
$$ 
1-Q(a|n,x) = {1 \over D}  
\int_a^\infty e^{-\theta}\prod_{i=1}^n [b + \theta r (x_i)]d\theta 
$$ 
\begin{equation} 
= {1 \over D}\sum_{m=0}^nb^{n-m}C_{n,m}\left({n \atop m} 
\right)\int_a^\infty \theta^m e^{-\theta}d\theta.  
\end{equation} 
We use the reasoning described after Equation(8) to evaluate this integral. 
$$ 
1-Q(a|n,x) = {1 \over D}\sum_{m=0}^n b^{n-m}C_{n,m}{n! \over 
m!(n-m)!}m!e^{-a}\sum_{l=0}^m{a^l \over l! } 
$$ 
\begin{equation} 
={e^{-a}\sum_{i=0}^n {b^{n-i}C_{n,i}\over (n-i)!}\sum_{l=0}^i  
{a^l \over l!} \over \sum_{m=0}^n {b^{n-m} \over (n-m)!}C_{n,m}}. 
\end{equation} 
We can use either Equation (19), recognizing that the integral is an 
incomplete gamma function, or use Equation (20) to find $Q(a|n,x)$. The 
limits can then be found by iterations as follows. 
 
\begin{enumerate} 
\item  First solve the equation $Q(z|n,x)=1-\epsilon$ for $z$. 
This is straightforward, since $Q(z|n,x)$ is increasing in $z$. 
\item  If $q(z|n,x)\le q(0|n,x)$, then $\ell=0$ and $u=z$. 
\item  If $q(0|n,x)< q(z|n,x)$, solve the equations 
\end{enumerate} 
$$q(y|n,x) = c,$$ 
$$q(z|n,x) = c,$$ 
$$Q(z|n,x) -Q(y|n,x) = 1-\epsilon$$ 
for $c,y,z$ by a double iteration.  In solving these equations, start 
with $c=(c_{min}+c_{max})/2$, where $c_{min}=q(0|n,x)$ and $c_{max}$ is 
the maximum value of $q(\theta|n,x)$, corresponding to 
$\theta=\theta_{max}$.  Solve the first two equations for $y<\theta_{max}<z$. 
This is straightforward, since $q(\theta|n,x)$ is increasing for 
$\theta\le \theta_{max}$ and decreasing for $\theta>\theta_{max}$. 
Then replace $c_{max}$ (respectively, $c_{min}$) by $c$, accordingly 
as $Q(z|n,x)-Q(y|n,x)<1-\epsilon$ (respectively, $>1-\epsilon$). 
 
Note that the iteration procedure for the Poisson example without additional 
parameters measured is just a special case of this iteration procedure. 
 
\section{Concluding Remarks} 
 
We have proposed methods for setting credible/confidence intervals for 
the {\it means} $\theta$ of a Poisson variable, observed in the presence of 
background, and of a normal distribution, when $\theta$ is known to be 
non-negative.  In the process, we have made specific choices for the 
prior distributions and loss structure.  The (uniform) prior 
distributions were chosen for mathematical tractability and agreement 
between the Bayesian credible level and conventional confidence level; 
and one of the main findings is that such agreement is possible.  The 
intervals have an optimality property: they minimize the length of 
credible intervals in the $\theta$ scale for the uniform prior. For
both problems, the mean provides a physically meaningful and  
mathematically tractable metric.  Our procedures depend on the choice 
of metric, prior, and loss structure as follows:  if 
the interval for $\theta$ is $\ell\le\theta\le u$ and the metric were 
changed, to $\tau= 1/\theta$ say, then a valid interval for $\tau$ is 
$1/u\le\tau\le 1/\ell$.  This has the same frequentist coverage as the 
original interval and is an exact credible interval for the induced 
prior, $d\tau/\tau^2$.  The optimality is lost, however.  The 
transformed interval does not minimize length in the $\tau$-scale.
The situation for our method is similar to that in the Cramer-Rao
limit, Equation 13.10 of Roe \cite{roebook}.  This very useful bound
also depends on the metric.   
 
The derivations of the ad hoc modifications in Section 5 mix Bayesian
and frequentist reasoning, and this may seem inelegant.  We believe
that any inelegance is mitigated by the minor nature of the changes
and the resulting good properties of the procedure from both the
Bayesian and frequentist viewpoints.

In the case of a Poisson distribution without auxiliary variables, 
there is a close connection between our method and $CL_s$: the $CL_s$ 
procedure is mathematically equivalent to an upper credible bound with 
a uniform prior distribution.  Read  \cite{readone}  
calls this a coincidence.  We think that there is a deeper  
connection \cite{hwangone}. 
The optimality claimed for $CL_{b+s}$ in Read's work is optimality for 
testing background only versus background plus a specified signal. 
Our procedures are designed to produce intervals of shortest length, 
not most powerful tests.  That these two criteria can lead to 
different procedures was noted in the statistical literature by  
Pratt \cite{prattone}.  
 
This research is supported by the National Security Agency under 
MDA904-99-1-004 and the National Science Foundation under  
PHY-9725921.

\appendix 
\section*{Proofs of Coverage and Optimality Theorems} 
\subsection{Continuous Example} 
 
Recall that if $x=\theta+\Delta$ is observed, then necessarily 
$\Delta\le x$ and that $\theta\in [\ell(x),u(x)]$ iff 
$-d(x)\le\Delta\le a(x)$, where $d(x)=u(x)-x$ and 
$a(x)=x-\ell(x)=min[d(x),x]$.  Let $x'=\theta+\Delta'$ be an 
independent copy of $x$. 
 
Theorem: {\it For every $x,\ Prob_\theta[-d(x)\le\Delta'\le 
a(x)|\Delta'\le x]=1-\epsilon$, and $[\ell(x),u(x)]$ is the shortest 
interval with this property.}  The first statement says that the 
conditional coverage of $x-\theta$ is exact. 
 
Proof:  The result follows from 
\begin{equation} 
Prob_\theta[\Delta'\le z|\Delta'\le x]= { \Phi[min(x,z)] \over 
\Phi(x) } = Prob[x-\theta\le z|x]. 
\end{equation} 
In words, the conditional frequentist distribution of $x'-\theta$ is 
the same as the posterior distribution of $x-\theta$. 
It follows that 
\begin{eqnarray*} 
Prob_\theta[-d(x)\le\Delta'\le a(x)|\Delta'\le x] &=& Prob[x-u(x)\le 
x-\theta\le min(x,d(x))|x] \\ 
	&=& Prob[\ell(x)\le\theta\le u(x)|x] =1-\epsilon,  
\end{eqnarray*} 
where the last equality follows from the definitions of $\ell(x)$ and 
$u(x)$.  Next, if $[L(x),U(x)]$ is any interval for which 
$Prob_\theta[-D(x)\le\Delta'\le A(x)|\Delta'\le x] = 1-\epsilon$, 
where $D(x) = U(x) -x$ and $A(x) = x-L(x)$, then 
$$ 
Prob[L(x)\le\theta\le U(x)|x]=Prob[-D(x)\le x-\theta\le A(x)|x]=1-\epsilon 
$$ 
and, therefore, $U(x)-L(x)\ge u(x)-\ell(x)$, since the Bayesian limits 
were chosen to minimize the length of the interval. 
   
\rightline{$\diamondsuit$} 
 
To put this in 
more conventional terms, let the interval $C(x,x') 
=[x'-a(x),x'+d(x)]$.  Then  
\begin{equation} 
Prob_\theta [C(x,x')\ni\theta |\Delta'\le x] = 1-\epsilon,  
\end{equation} 
where $x'$ is the random variable. 
 
Theorem: {\it A lower limit for the conventional coverage is 
$(1-\epsilon)/(1+\epsilon)$.} 
 
Proof:  Recall that $x_\ell$ and $x_u$ were the values of $x$ for a 
given $\theta$ meeting the upper and lower Bayesian limits, so 
$\theta = u(x_\ell) = x_\ell + d(x_\ell)$ and $\theta = \ell(x_u) = x_u - 
min[d(x_u),x_u]$.  The unconditional frequentist coverage probability 
is  
$$ 
Prob_\theta[\ell(x)\le \theta\le u(x)] = Prob_\theta[x_\ell\le x\le x_u] = 
\Phi(x_u-\theta) -\Phi(x_\ell - \theta). 
$$   
Recall the definition of $x_0$ and observe that $x_u\ge x_0$ for all 
$\theta>0$, since $\theta=\ell(x_u)$ and $\ell(x)=0$ for $x\le x_0$. 
First suppose that $x_\ell \le x_0$.  Then  
$\Phi(x_\ell) \le \Phi(x_0)\le \Phi(x_u)$
and $\Phi(x_0) = 1/(1+\epsilon)$ by (6).  Also, by (4) and (5), 
$\Phi(x_\ell-\theta) = \Phi[-d(x_\ell)] = 
1-\Phi[d(x_\ell)]=\epsilon\Phi(x_\ell),\ \Phi(x_u-\theta)=\Phi[d(x_u)] 
= { 1 \over 2} + { 1 \over 2}(1-\epsilon)\Phi(x_u),$ and   
\begin{eqnarray*} 
`\Phi(x_u-\theta) - \Phi(x_\ell-\theta) &=& {1 \over 2} + {1 \over
2}(1-\epsilon) \Phi(x_u) -\epsilon\Phi(x_\ell) \\  
&\ge& {1 \over 2} + {1 \over 2}{ 1-\epsilon \over 1+
\epsilon} -  
{\epsilon \over 1+ \epsilon } \\ 
		&\ge& {1-\epsilon \over 1+\epsilon}.  
\end{eqnarray*} 
Next suppose $x_\ell > x_0$.  Then 
$\Phi(x_u)\ge\Phi(x_\ell)\ge\Phi(x_0)$,  
$\Phi(x_\ell-\theta)=\Phi[-d(x_\ell)]= 1-\left[{1 \over 2} + {1 \over 
2}\left(1-\epsilon\right)\Phi(x_\ell)\right]$, and 
\begin{eqnarray*} 
	\Phi(x_u-\theta) -\Phi(x_\ell -\theta) &=& {1 \over 2} + {1 \over 
2}(1-\epsilon )\Phi(x_u)-1 +\left[ {1 \over 2} + {1 \over 
2}(1-\epsilon)\Phi(x_\ell) \right] \\ 
&=& {1 \over 2 }(1-\epsilon)[\Phi(x_\ell) +
\Phi(x_u)] \\  
&\ge& {1-\epsilon \over 1+ \epsilon}. 
\end{eqnarray*} 
   
\rightline{$\diamondsuit$}

Theorem: {\it For the modified procedure a lower limit to the 
conventional coverage is given by ${ 1 \over 2} + { 1 \over 2} 
(1-\epsilon)\Phi(x_u) - 
min\left({1 \over 2},\Phi(x_l)\right)\epsilon$.} 
 
Proof:  For the modified procedure, the upper limit is defined as 
$$ 
u'(x) = \max\left[u(x),x+\Phi^{-1}\left(1-{1 \over 
2}\epsilon\right)\right]. 
$$ 
We set $x'_\ell$ to be the $x$ corresponding to $u'(x) =\theta$.  Then 
$x'_\ell=min[x_\ell,\theta-\Phi^{-1}(1-{ 1 \over 2}\epsilon)]$, and 
$x'_\ell-\theta \le -\Phi^{-1}(1-\epsilon/2)$.   
Observe that $u'(x) = u(x)$ for $x\le 0$ and $u'(x) = 
\Phi^{-1}(1-{ 1 \over 2}\epsilon)$ for $x>0$.  For if $x\le 0$, then $u(x)-x = 
d(x) =\Phi^{-1}[1-\epsilon\Phi(x)]>\Phi^{-1}(1-{ 1 \over 2}\epsilon)$ using 
Equation (5).  Similarly, if $0<x<x_0$, then 
$u(x)-x=d(x)=\Phi^{-1}[1-\epsilon\Phi(x)]<\Phi^{-1}(1-{ 1 \over 2}\epsilon)$; 
and if $x_0<x<\infty$, then $u(x)-x=d(x) = 
\Phi^{-1}[{ 1 \over 2}-{ 1 \over 2}(1-\epsilon)\Phi(x)]<\Phi^{-1}(1-{ 
1 \over 2}\epsilon)$.  It  
follows that $x'_\ell = x_\ell$ for $x_\ell<0$ and 
$x'_\ell=\theta-\Phi^{-1}(1-{ 1 \over 2}\epsilon)$ if $x_\ell>0$. 
 
So, if $x_\ell>0$, then $x'_\ell -\theta = -\Phi^{-1}(1-{1 \over 
2}\epsilon) = \Phi^{-1}({1 \over 2}\epsilon)$ and, therefore, 
$$ 
\Phi(x'_\ell-\theta) 
=\Phi[\Phi^{-1}( {1 \over 2}\epsilon)] = {1 \over 2}\epsilon. 
$$ 
If $x_\ell<0(<x_0)$, then 
$$ 
\Phi(x'_\ell-\theta)\equiv\Phi(x_\ell-\theta) = \epsilon\Phi(x_\ell).$$ 
Hence  
\begin{equation} 
\Phi(x'_\ell-\theta) = min\left[{1 \over 2},\Phi(x_\ell)\right]\epsilon,  
\end{equation} 
and the theorem follows from $\Phi(x_u-\theta)-\Phi(x'_\ell-\theta) = 
{ 1 \over 2} +{ 1 \over 2} (1-\epsilon)\Phi(x_u) -\Phi(x'_\ell-\theta)$. 
   
\rightline{$\diamondsuit$} 
 
\subsection{Discrete(Poisson) Example} 
 
{\it The Bayesian credible intervals are (nearly) exact 
conditional confidence intervals (except for discreteness) if we 
convert to a scale that is natural for conditional confidence.} 
 
	To see this, let  
$$ 
H_a(y) =\int_0^y {x^a \over \Gamma(a+1)}e^{-x}dx, 
$$ 
where $\Gamma$ is the Gamma function.  Then, using reasoning similar 
to that following Equation (8), 
$H_n(\mu) = 1 -P_\mu(n)$.  From Equation (9), the posterior
distribution function of $\theta$ is 
\begin{eqnarray*} 
Prob(\theta\le\theta_0|n) \equiv Q(\theta_0|n) &=&
\int_0^{\theta_0} p_b(\theta|n)
d\theta \\  
&=& {H_n(b+\theta)-H_n(b) \over P_b(n)}.  
\end{eqnarray*} 
Fix $b$ and let $K_{\theta,n}(m)$ be the conditional probability of at 
most $m$ events from the conditional sample space, given 
at most $n$ background events in the experiment, {\it i.e.},\break\hfill  
$K_{\theta,n}(m)=Prob_\theta(\le  
m{\rm\ events\ obtained}|n_{background}\le n)$.  Then 
  
$$ 
K_{\theta,n}(m) =  {P_{\theta+b}(m) \over P_b(n)} 
$$ 
for $m\le n$, and 
$$ 
K_{\theta,n}(m) = {1 \over P_b(n)}\sum_{k=0}^n{e^{-b} \over 
k!}b^kP_\theta(m-k) 
$$ 
for $m>n$.  Letting $M$ be the random count obtained in a particular 
experiment 
\begin{equation} 
Prob_\theta[K_{\theta,n}(M)\le y|n_{background}\le n]\approx y
\label{eq:approx} 
\end{equation} 
for $0\le y\le 1$, where the approximation arises because $m$ is 
discrete.  Specifically, (\ref{eq:approx}) is exact for $y$ of the form  
$y = K_{\theta,n}(m)$, since $Prob_\theta[K_{\theta,n}(M)\le K_{\theta,n}(m)| 
n_{background}\le n] =Prob_\theta[M\le m|n_{background}\le n]=
K_{\theta,n}(m)$ 
for all $m= 0,1,2,\cdots$. 
 
Clearly, $\ell_n\le\theta\le u_n$ iff $b+\ell_n\le b+\theta\le b+u_n$ 
iff $H_n(b+\ell_n)\le H_n(b+\theta)\le H_n(b+u_n)$.  Thus, 
$\ell_n\le\theta\le u_n$ iff $1-H_n(b+u_n)\le P_{\theta+b}(n)\le 1-
H_n(b+\ell_n)$, or  
equivalently,  
$$  
{1 - H_n(b+u_n) \over P_b(n)} \le K_{\theta,n}(n) \le { 1- 
H_n(b+\ell_n) \over P_b(n)}. 
$$ 
Using (\ref{eq:approx}), the conditional probability of the latter
event given  
$n_{background} \leq n$ is approximately 
\begin{equation} 
{H_n(b+u_n)-H_n(b+\ell_n) \over P_b(n)} = 1-\epsilon,
\label{eq:approx2}  
\end{equation} 
where the last equality follows from the definitions of $\ell_n$ and 
$u_n$.

To appreciate (\ref{eq:approx2}), it is instructive to use a different 
approach.  Suppose that we knew apriori that $n_{background}\le n$.  Then the  
mass function of $M$ is then  
$$ 
Prob_\theta(M=m|n_{background}\le n) \equiv 
k_{\theta,n}(m) = K_{\theta,n}(m) -  K_{\theta,n}(m-1). 
$$   
To form 
confidence intervals in this model, we might construct tests that the 
parameter has a given value and then invert this family of tests. 
This requires finding $n_\ell$ and $n_u$ for which 
$Prob_\theta(n_\ell\le M\le  
n_u|n_{background}\le n)\ge 1-\epsilon$ for each $\theta$, where 
$n_\ell$ and $n_u$ are functions of $\theta$ and $n$.  This implies 
and is nearly equivalent to 
\begin{equation} 
Prob_\theta[K_{\theta,n}(n_\ell)\le K_{\theta,n}(M) \le 
K_{\theta,n}(n_u)|n_{background}\le n]\ge 1-\epsilon. \label{eq:cndcnf} 
\end{equation} 
Except for discreteness, the left side of (\ref{eq:cndcnf}) is 
$K_{\theta,n}(n_u) -K_{\theta,n}(n_\ell)$.  The Bayesian credible 
intervals implicitly determine values of $n_\ell$ and $n_u$, namely 
$$ 
K_{\theta,n}(n_\ell) = { 1- H_n(b+u_n) \over P_b(n)} 
$$ 
$$ 
K_{\theta,n}(n_u) ={ 1- H_n(b+\ell_n) \over P_b(n)}. 
$$ 
The relation (\ref{eq:approx2}) shows that these values
nearly satisfy 
(\ref{eq:cndcnf}), except for discreteness.

\begin{figure} 
\epsfig{file=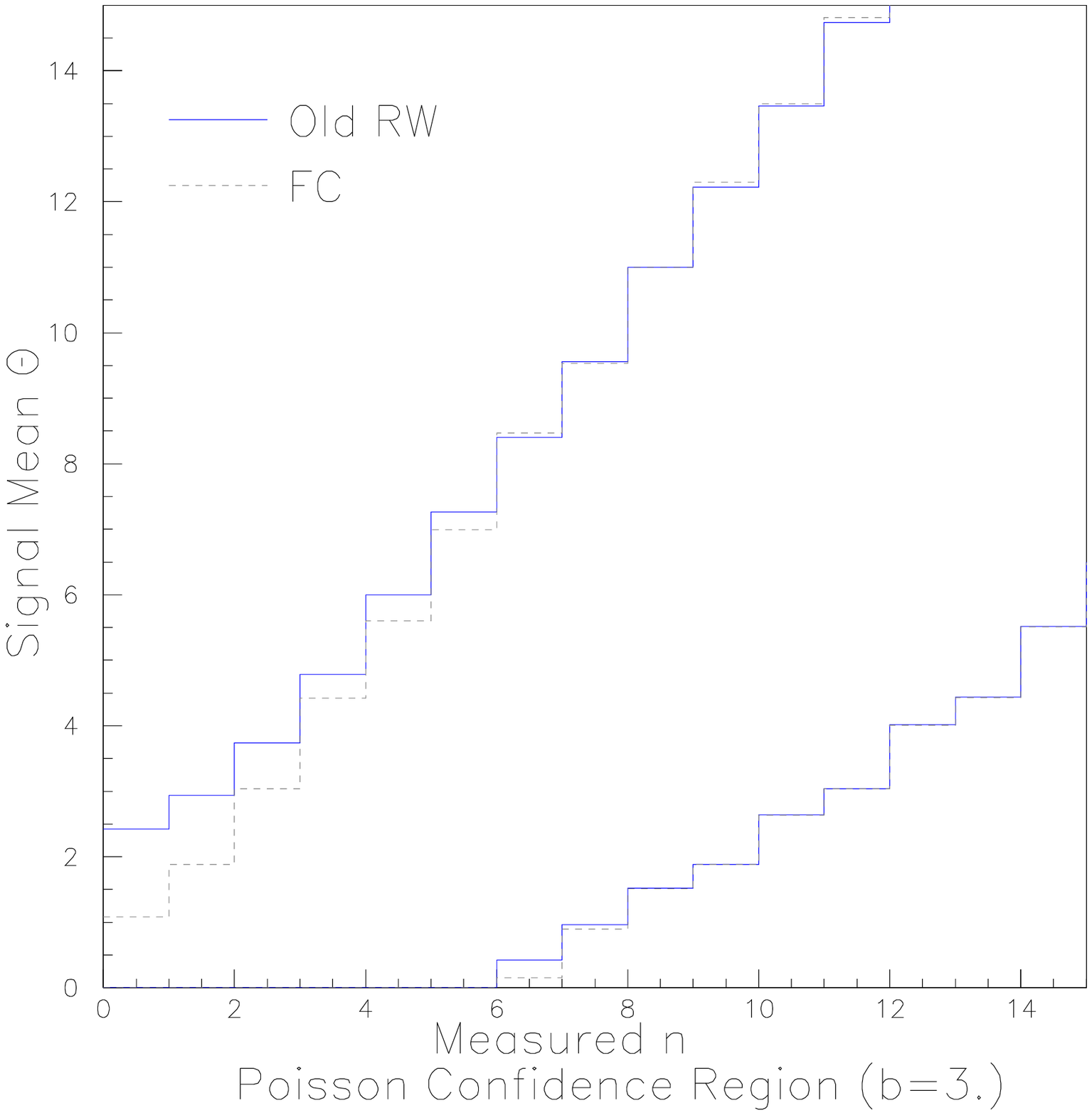,height=5in} 
\bigskip 
\caption{The 90\% C.L. belt for a Poisson probability with $b=3$,  
using the old R-W procedure (solid line) and the Feldman-Cousins unified 
procedure (dashed line).} 
\end{figure} 
 
\newpage 
 
\begin{figure} 
\epsfig{file=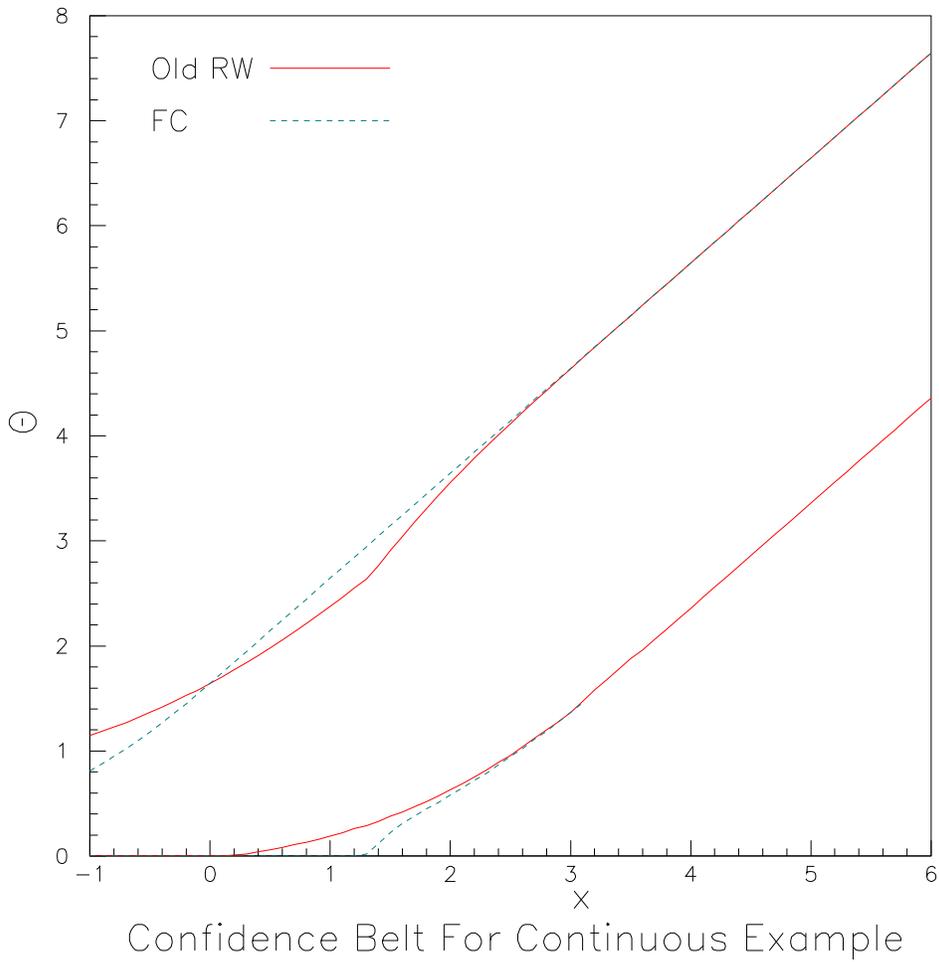,height=5in} 
\bigskip 
\caption{The 90\% C.L. belt for the continuous probability example,  
using the old R-W procedure (solid line) and the  
Feldman-Cousins unified procedure (dashed line).} 
\end{figure} 
 
\newpage 
 
\begin{figure} 
\epsfig{file=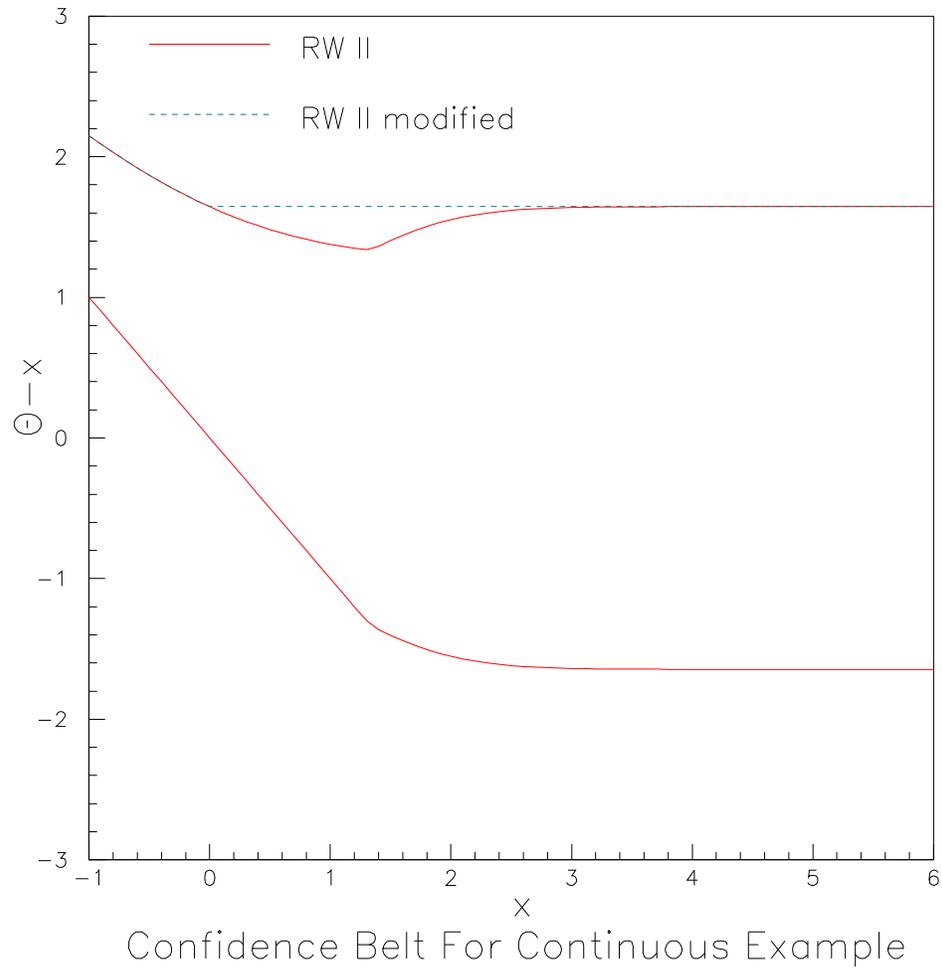,height=5in} 
\bigskip 
\caption{The 90\% C.L. belt plotted as $\theta -x$ vs 
$x$ using the 
Bayesian procedure and a conservative modification (dashed) for the continuous 
example.} 
\end{figure}

\newpage 
 
\begin{figure} 
\epsfig{file=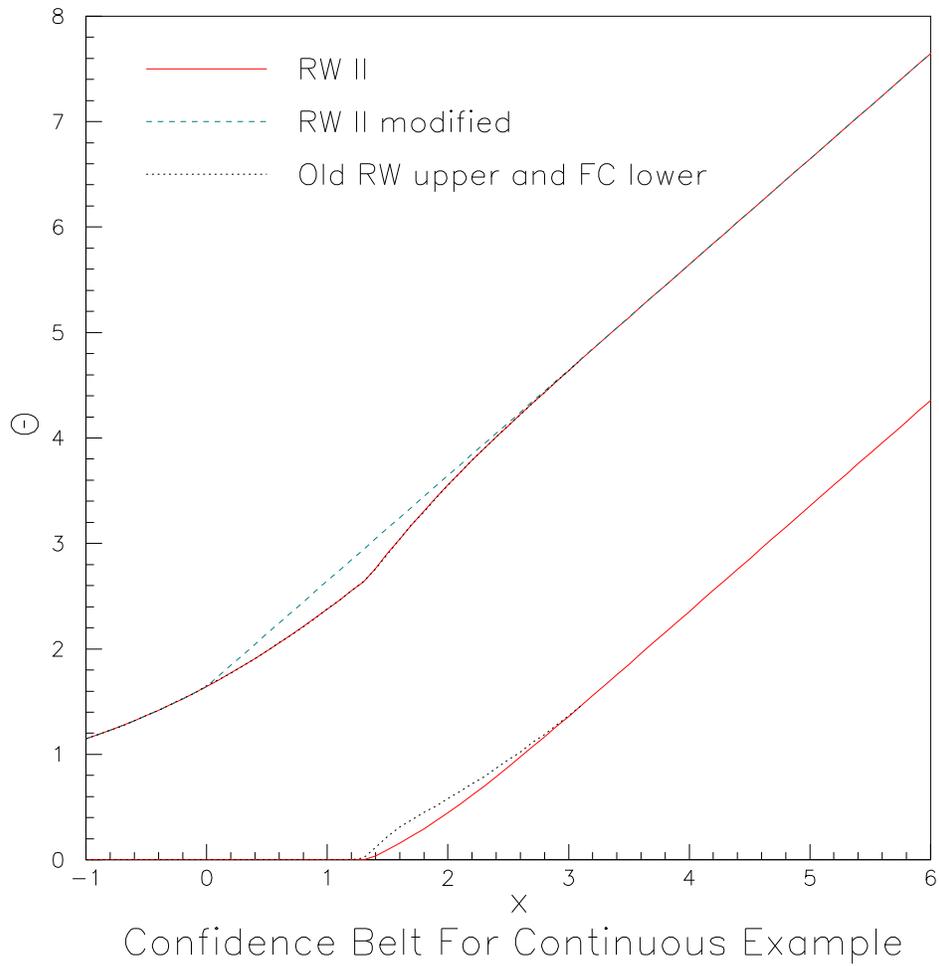,height=5in} 
\bigskip 
\caption{The 90\% C.L. belt using the 
Bayesian procedure (solid line) and the conservative modification 
(dashed line) for  
the continuous 
example.  The old R-W upper limit and the F-C unified method lower 
limit are shown as dotted lines.} 
\end{figure} 
 
\newpage

\begin{figure} 
\epsfig{file=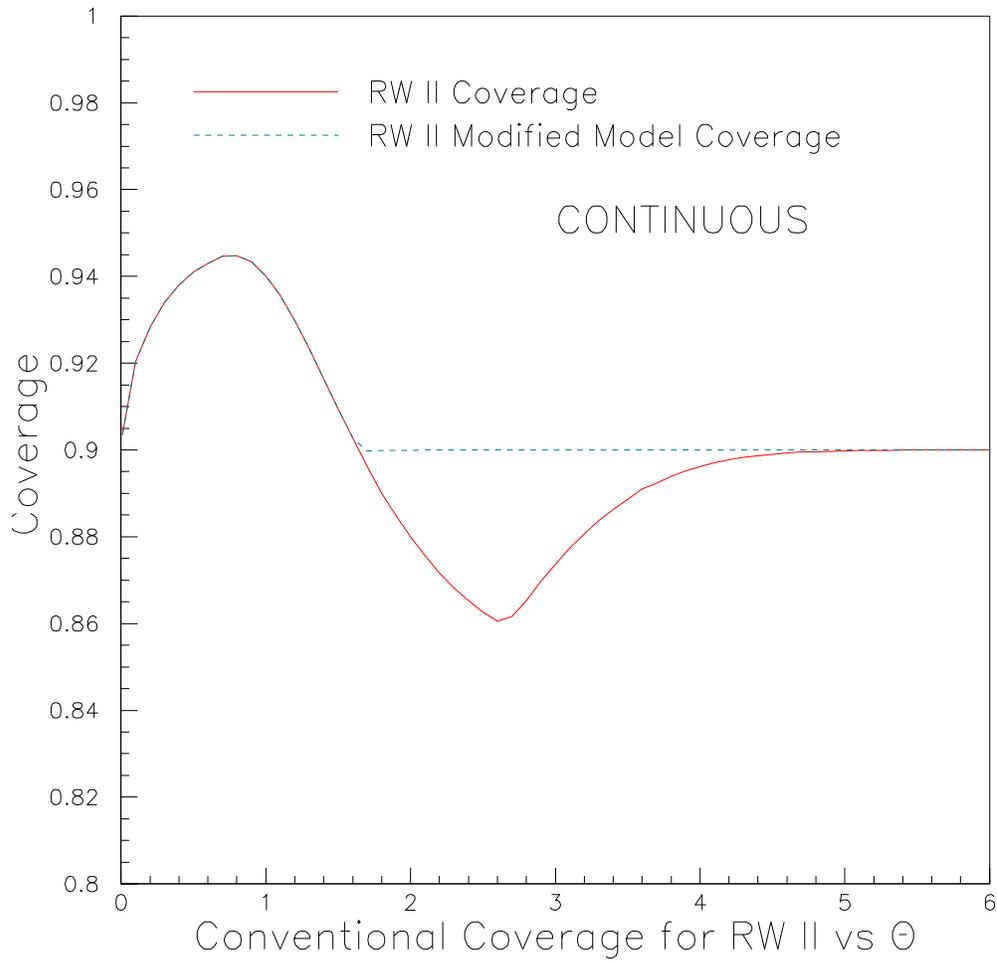,height=5in} 
\bigskip 
\caption{The conventional frequentist coverage for a Bayesian 
90\% C.L. belt for the Bayesian continuous model (solid line) and the  
conservative modification (dashed line).} 
\end{figure} 
 
\newpage 
 
\begin{figure} 
\epsfig{file=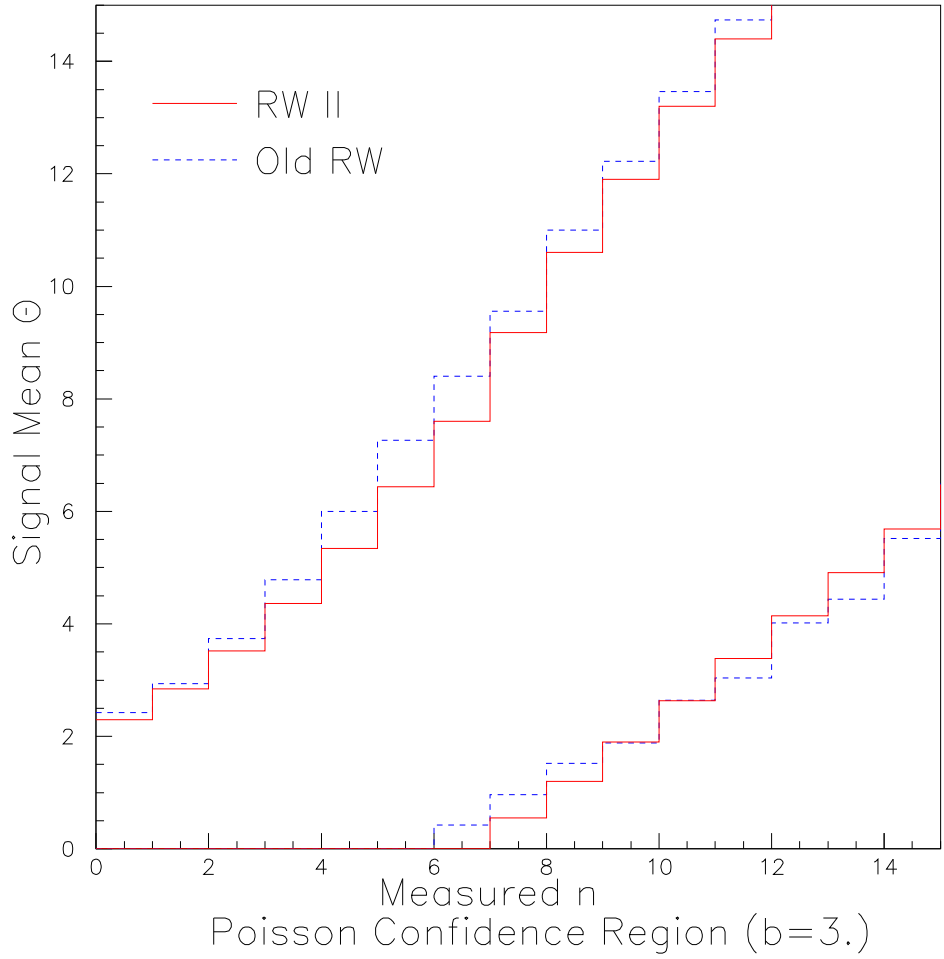,height=5in} 
\bigskip 
\caption{The 90\% C.L. belt for the Poisson distribution with $b=3$ using the 
Bayesian procedure and the 
old R-W procedure (dashed line).} 
\end{figure} 
 
\newpage 
 
\begin{figure} 
\epsfig{file=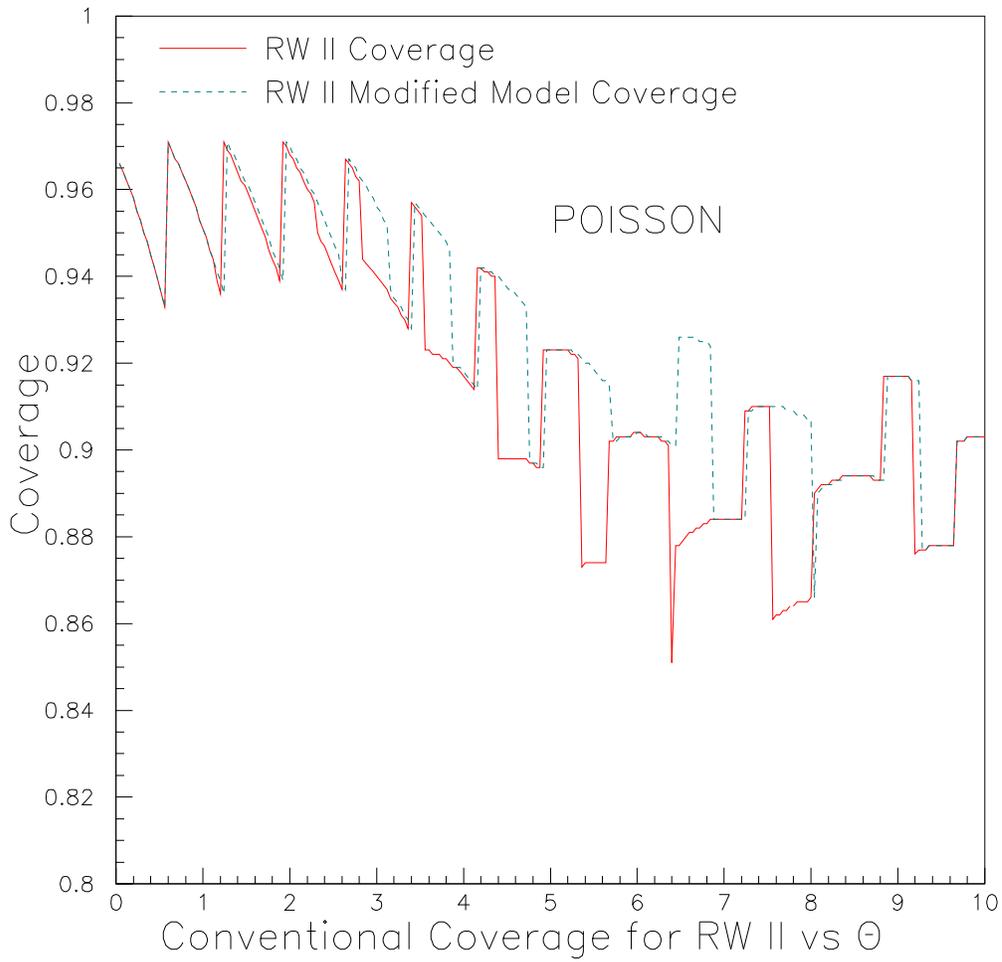,height=5in} 
\bigskip 
\caption{The conventional frequentist coverage obtained by this Bayesian 
procedure (solid line) and by the conservative modification  
(dashed line) for $b=3$.} 
\end{figure} 
 
\newpage 
 
\begin{figure} 
\epsfig{file=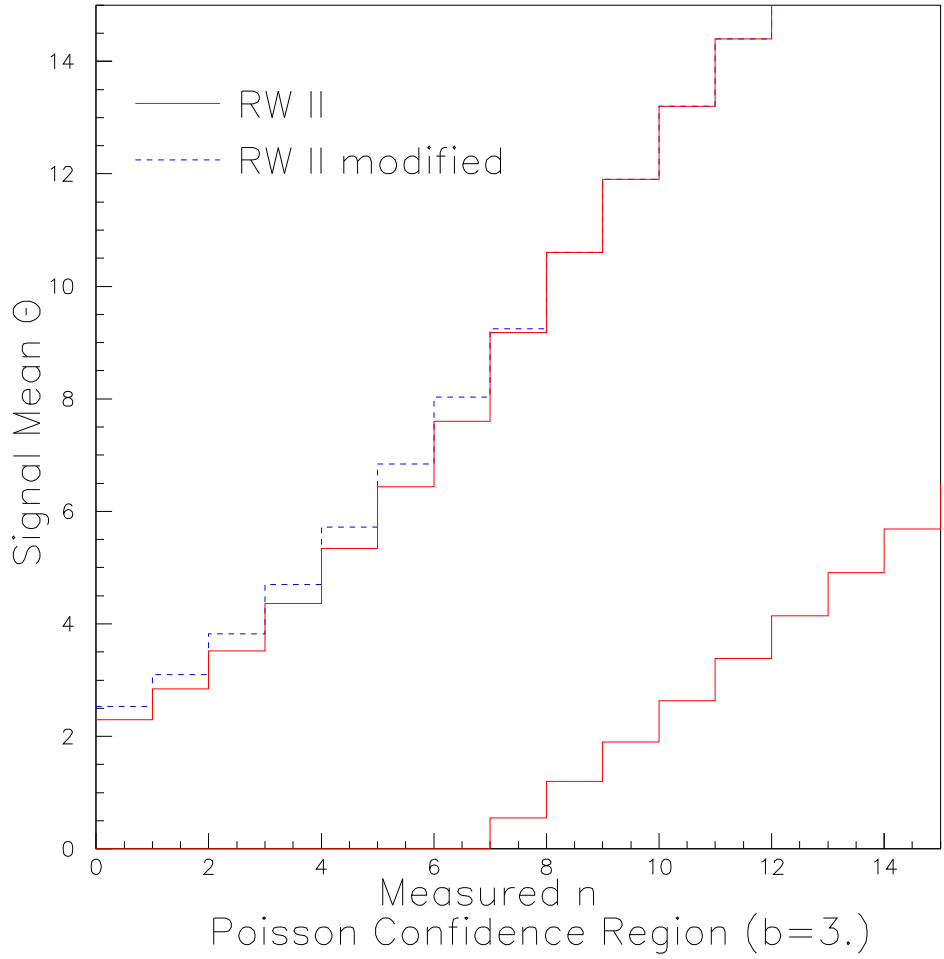,height=5in} 
\bigskip 
\caption{The 90\% C.L. belt for the Poisson distribution with $b=3$ using the 
Bayesian procedure (solid line) and the conservative modification  
(dashed line).} 
\end{figure}

\end{document}